\documentclass[12pt]{article}
\usepackage{amsfonts,amssymb,epsfig}
\usepackage[latin1]{inputenc}
\usepackage[english]{babel}
\usepackage{hyperref}
\usepackage{color}
\newtheorem{thm}{Th\'eor\`eme}[section]
\newtheorem{cor}[thm]{Corollaire}
\newtheorem{lem}[thm]{Lemme}
\newtheorem{pro}[thm]{Proposition}
\newtheorem{dfn}[thm]{D\'efinition}
\newtheorem{rmk}[thm]{Remark}
\newtheorem{expl}[thm]{Exemple}

\def\dessous#1\sous#2{\mathrel{\mathop{\kern0pt#2}\limits_{#1}}}

\newcommand{\R}{\mathbb R}
\newcommand{\C}{\mathbb C}
\newcommand{\N}{\mathbb N}

\newcommand{\1}{1 \! \! {\rm I}}

\newcommand{\beq}{\begin{eqnarray}}
\newcommand{\eeq}{\end{eqnarray}}
\newcommand{\bpro}{\begin{pro}}
\newcommand{\epro}{\end{pro}}
\newcommand{\blem}{\begin{lem}}
\newcommand{\elem}{\end{lem}}
\newcommand{\bdfn}{\begin{dfn}}
\newcommand{\edfn}{\end{dfn}}
\newcommand{\bcor}{\begin{cor}}
\newcommand{\ecor}{\end{cor}}
\newcommand{\bthm}{\begin{thm}}
\newcommand{\ethm}{\end{thm}}
\newcommand{\bex}{\begin{expl}}
\newcommand{\eex}{\end{expl}}
\newcommand{\brmk}{\begin{rmk}}
\newcommand{\ermk}{\end{rmk}}
\newcommand{\benum}{\begin{enumerate}}
\newcommand{\eenum}{\end{enumerate}}
\newcommand{\bitem}{\begin{itemize}}
\newcommand{\eitem}{\end{itemize}}

\begin{document}
\begin{center}
{\Large\bf {Coherent states for a system of an electron moving on  plane}}\\
\vspace{0.5cm}
Isiaka Aremua$^{1,2}$ and Laure Gouba$^{3}$ \\
$^{1}${\em
Universit\'{e} de Lom\'{e} (UL), Facult\'{e} Des Sciences  (FDS), D\'{e}partement de Physique} \\
{\em Laboratoire de Physique des Mat\'eriaux et des Composants \`a Semi-Conducteurs}\\
{\em Universit\'{e} de Lom\'{e} (UL), 01 B.P. 1515  Lom\'{e} 01, Togo.}\\  
 $^{2}$International Chair of Mathematical Physics
 and Applications. \\
 {\em ICMPA-UNESCO Chair,  University of Abomey-Calavi}\\
 {\em 072 B.P. 50 Cotonou, Republic of Benin.\\
 E-mail: claudisak@gmail.com}\\
$^{3}${\em
The Abdus Salam International Centre for Theoretical Physics (ICTP),\\
Strada Costiera 11, I-34151 Trieste Italy.\\
 E-mail: lgouba@ictp.it}

\vspace{1.0cm}

\today

\begin{abstract}
\noindent
In this paper, we construct the coherent states for a system of an electron moving on plane in uniform external magnetic and electric fields. These coherent states are built in the context of both discrete and continuous spectra and satisfy the Gazeau-Klauder coherent states properties \cite{gazeau-klauder}. 
\end{abstract}

\end{center}
\setcounter{footnote}{0}

\section{Introduction}

The system of charged quantum particles interacting with a constant magnetic field continues to attract intensive studies and is without a doubt one of the most investigated quantum systems, mainly motivated by condensed matter physics and quantum optics. A review article devoted to  this  quantum system and its related different kind  of coherent states (CSs) was recently  elaborated by Dodonov (see \cite{dodonov} and the complete reference list therein).

The concept of what is now called coherent states has been of great interest to the scientific community since the work of 
Schr\"{o}dinger in 1926 \cite{schroedinger} on  the quantum harmonic { oscillator (HO), where he introduced} a  specific
 quantum state that has dynamical behavior that is most similar to that of the classical HO. The conditions any state must fulfill to be coherent were elaborated by Klauder as follows:  continuity in complex label, normalization, non orthogonality, unity operator resolution with unique positive weight function of the integration measure, temporal stability and action identity \cite{klauderCS}. More details on the CSs and their different generalizations can be found in the literature \cite{klauder-skagerstam}-\cite{ali-antoine-gazeau}, the list is not of course exhaustive.

In his study \cite{landau}, Landau found that the system of
electronic motion in a static uniform magnetic field can be assimilated in two dimensions to a harmonic oscillator, with an energy structure of equidistant discrete levels,
with a distance $\hbar \omega_c$ ($\omega_c$ is the
cyclotron frequency), each level being highly
degenerate. Such a system, more often named Landau model, also provides a natural description for other well known significant phenomena, the so-called integer and fractional quantum Hall effects. {In these last years, in the search of understanding  the main features of the fractional quantum Hall effect (FQHE) \cite{pasquier, prange-girvin}}, many efforts  have been done in the literature to find a wave function which minimizes the energy of a two-dimensional system of electrons subjected to a strong constant magnetic field applied perpendicularly to the sample, independently of the electron density. In \cite{bagarello}, a system of electrons, essentially a two-dimensional crystal, has been considered. Besides, the wave function introduced has been modified to lower the energy in order to explain the experimental data. From an appropriate quantization of the classical variables of the system Hamiltonian, Bagarello  {\it et al} (see \cite{bagarello}, \cite{antoine-bagarello} and references therein), have modified the
single electron wave function in view of the study of localization properties. The similar quantization has been also used  to investigate the Bohm-Aharonov effect (\cite{harms-micu, omer-jellal} and references therein)  emphasizing the fact that it is not the electric and the magnetic fields but the electromagnetic potentials which are the fundamental quantities in quantum mechanics.

In a previous work \cite{ab1}, a connection has been established between quantum Hall effect and  vector coherent states (VCSs) \cite{thirulogasanthar-ali, ali-englis-gazeau} 
by applying the various construction methods developed in the
literature. In the same way, the motion of an electron in a noncommutative $xy$ plane, in a constant magnetic field background coupled with a harmonic potential was examined with the relevant VCSs constructed and discussed 
 \cite{aremuajnmp}. The Barut-Girardello CSs
 have been built  for Landau levels of a gas of spinless charged particles, subject to a perpendicular magnetic field confined in a harmonic potential with thermodynamical and the statistical properties have been  investigated \cite{aremuaromp}. See also \cite{dodonov} and references quoted therein. Recently \cite{aremuaastp}, from a  matrix (operator) formulation of the Landau problem and the corresponding Hilbert space, an  analysis
of various VCSs extended to diagonal matrix domains has been performed on the basis of Landau levels.

The construction of CSs for continuous spectrum was first proposed for the Gazeau-Klauder CSs in \cite{gazeau-klauder} and later in \cite{inomata, klauder2, popov}. 
In the present work,  we follow  the method developed in  \cite{gazeau-klauder},  by considering Landau levels,  to built various classes of  CSs as in \cite{ab1, gazeau-novaes, gouba} arising from  physical Hamiltonian describing a charged particle in an electromagnetic field, by introducing  additional parameters useful for handling  discrete and continuous spectra of the Hamiltonian. The eigenvalue problem is presented and the quantum Hamiltonian spectra provided in the two possible orientations of the magnetic field by considering the infinite degeneracies of the Landau levels. The CSs are constructed with relevant properties discussed for both continuous and discrete spectra, and for purely discrete spectrum. 

The  paper is organized as follows. In section \ref{sec2},  we revisit  the   model of electron moving on plane where 
 the eigenvalue problems are explicitely set  and solved.
The position and momentum operators, satisfying canonical commutation relations, established for the considered Hamiltonians are  also defined. Section \ref{sec3} is
devoted to the construction of CSs for the quantum 
Hamiltonian possessing both continuous and discrete spectra by following the  method developped in \cite{gazeau-klauder, ab1}. Concluding remarks are given in Section \ref{sec4}.

 \section{{Electron moving on plane revisited}}\label{sec2}
 
 In this section, we revisit the system of an electron moving on plane as in \cite{omer-jellal}, where we consider different scenarios for the symmetric gauge and the scalar potential.
 
 Consider an electron moving on the plane $xy$ in the uniform external
electric field  $\overrightarrow{E} =-\overrightarrow{\nabla}\Phi(x,y)$ and the uniform external magnetic field $\overrightarrow{B}$ which is perpendicular to the plane  described by the Hamiltonian

\beq{\label{es0}}
H = \frac{1}{2m}\left(\overrightarrow{p} +
\frac{e}{c}\overrightarrow{A}\right)^{2} - e\Phi.
\eeq

\subsection{Case of the symmetric gauge} 

\beq{\label{es01}}
\overrightarrow{A} = \left(\frac{B}{2}y, -\frac{B}{2}x \right). 
\eeq

Experimentally, the electric field $\overrightarrow{E}$ is oriented according to one of the two possible directions of the plane. Suppose the scalar potential is defined as \beq{\label{es02}} \Phi(x,y) = -Ey. \eeq
Substituting the relations (\ref{es01}) and  (\ref{es02}) in
(\ref{es0}), the corresponding  classical Hamiltonian, denoted by $H_{1}$, reads 

\beq{\label{es1}} 
H_{1}(x,y,p_x,p_y)   =
\frac{1}{2m}\left[\left(p_{x} + \frac{eB}{2c}y\right)^{2} +
\left(p_{y} - \frac{eB}{2c}x\right)^{2} \right] + eEy. 
\eeq 
A canonical quantization of this system is obtained by promoting the classical variables $x,y,p_x,p_y$, to the operators $X,Y,P_x,P_y$ which satisfy the nonvanishing canonical commutation relations
\beq
[X, P_x] = i\hbar = [Y, P_y].
\eeq
The Hamiltonian operator is derived from (\ref{es1}) as follows
\beq
\hat H_1(X,Y,P_x,P_y) = \frac{1}{2m}\left[\left(P_{x} + \frac{eB}{2c}Y\right)^{2} + \left(P_{y} - 
\frac{eB}{2c}X\right)^{2}\right] + eEY.
\eeq
In order to solve the eigenvalue problem
\beq
\hat H_1\Psi = \mathcal{E}\Psi, 
\eeq
it is convenient to perform the change of variables as below
\beq{\label{complxop}} 
Z = X + i Y, \;\;\;  P_{z} =
\frac{1}{2}(P_{x} - i P_{y}), 
\eeq
satisfying  the nonvanishing commutations relations
\beq
[Z, P_z] = i\hbar = [\bar Z, P_{\bar z}], 
\eeq
and to define two sets of annihilation and creation operators $b, b^{\dag}$  and $d, d^{\dag}$
given by 
\beq{\label{es2}} 
b = 2P_{\bar{z}} - i
\frac{eB}{2c}Z + \lambda, \qquad b^{\dag} = 2P_{z} + i
\frac{eB}{2c}\bar{Z} + \lambda, 
\eeq 
\beq{\label{es3}} d =
2P_{\bar{z}} + i \frac{eB}{2c}Z, \qquad  d^{\dag} = 2P_{z} -
i \frac{eB}{2c}\bar{Z}, 
\eeq
with $\lambda = \frac{mcE}{B}.$
These two sets of operators commute each other and satisfy the
following commutation relations \beq{\label{es4}} 
[b, b^{\dag}] = 2 m \hbar \omega_{c} {\1}, \quad
 [d^{^{\dag}}, d] = 2 m \hbar \omega_{c} {\1},
\eeq 
where $\omega_{c} = \frac{eB}{mc}$ is known as the cyclotron frequency and ${\1}$ is the unit operator. The Hamiltonian $\hat H_1$ can be then re-expressed as follows:
\beq\label{es7}
\hat H_1 = \frac{1}{4m}\left(b^{\dag}b + bb^{\dag}\right) - \frac{\lambda}{2m}\left(d^{\dag} + d\right) - \frac{\lambda^{2}}{2m}.
\eeq
In order to compute the eigenvalues $\mathcal{E}$ and eigenvectors $\Psi$, we split $\hat H_1$ in  (\ref{es7}) into two commuting parts in the following manner:
 \beq 
\hat H_{1} = \hat H_{1_{OSC}} - \hat T_1, 
\eeq
where  $\hat H_{1_{OSC}}$ denotes the harmonic oscillator part 
\beq
\hat H_{1_{OSC}} = \frac{1}{4m}(b^{\dag}b + bb^{\dag}),
\eeq while the part linear in $d$ and  $d^{\dag}$ is given by 
\beq
\label{tfunc1}
\hat T_1 =
\frac{\lambda}{2m}(d^{\dag} + d) + \frac{\lambda^{2}}{2m}. 
\eeq 
The annihilation and creation operators $b$ and $b^{\dag}$ can be also rewritten as follows:
 \beq b = \sqrt{2m\hbar \omega_{c}}b',\quad b^{\dag} = \sqrt{2m\hbar \omega_{c}}b'^{\dag},\quad [b',b'^{\dag}] = \1,
\eeq 
with \beq{\label{es11}} 
b' = \sqrt{\frac{2m\omega_{c}}{\hbar}}\left(\frac{
P_{\bar{z}}}{m\omega_{c}} - i \frac{Z}{4} +
\frac{\lambda}{2m\omega_{c}}\right), \quad
 b'^{\dag} = \sqrt{\frac{2m\omega_{c}}{\hbar}}\left(\frac{ P_{z}}{m\omega_{c}} + i \frac{\bar{Z}}{4} + \frac{\lambda}{2m\omega_{c}}\right). 
\eeq
Then, one has 
\beq 
b|0\rangle  = 0, \quad 
b|n\rangle = \sqrt{n}\sqrt{2m \omega_{c}
\hbar}|n-1\rangle,  \quad 
b^{\dag}|n\rangle = \sqrt{n+1}\sqrt{2m \omega_{c}
\hbar}|n+1\rangle
\eeq 
leading to 
\beq
|n+1\rangle =
\frac{1}{\sqrt{2m \omega_{c} \hbar (n+1)}}b^{\dag}|n\rangle,
\eeq
and, recurrently, to 
\beq{\label{es13}} 
\Phi_n \equiv |n\rangle =
\frac{1}{\sqrt{(2m \omega_{c} \hbar)^{n} n
!}}(b^{\dag})^{n}|0\rangle. 
\eeq 
The harmonic oscillator Hamiltonian
$\hat H_{1_{OSC}}$ reduces to 
\beq\label{oscil1}
\hat H_{1_{OSC}} = \frac{\hbar
\omega_{c}}{2}(2 N' + \1), \; N'= b'^\dag b'
\eeq 
with eigenvalues
$\mathcal E_{n_{1_{OSC}}}$ given by \beq \mathcal E_{n_{1_{OSC}}} = \hbar \omega_{c}(n + \frac{1}{2}), \quad n = 0, 1, 2, \ldots, \eeq corresponding to the eigenvectors  defined by (\ref{es13}).

The eigenvalue equation $\hat T_1\phi = \mathcal E \phi$ can be reduced to \beq \left(-i \frac{\partial}{\partial x} -
\frac{m \omega_{c}}{2 \hbar}y \right)\phi - \frac{m \mathcal
E}{\hbar \lambda}\phi = 0. \eeq 
Setting $\alpha = \frac{m \mathcal
E}{\hbar \lambda}$, it becomes  \beq -i
\frac{\partial}{\partial x}\phi = \left(\frac{m \omega_{c}}{2
\hbar}y + \alpha\right)\phi,
\eeq 
whose solution is readily found to
be \beq{\label{es15}} \phi_{\alpha} \equiv \phi_{\alpha}(x,y) =
e^{i (\alpha x + \frac{m \omega_{c}}{2 \hbar}xy)}, \;\;\;
\alpha \in \R. \eeq Then, the  eigenvalues of the operator $\hat T_1$,
corresponding to  eigenfuctions (\ref{es15}),   are given by
\beq{\label{es16}} \mathcal E_{\alpha} = \frac{\hbar
\lambda}{m}\alpha + \frac{\lambda^{2}}{2m}, \;\;\; \alpha \in \R,
\eeq indicating that this spectrum, labeled by $\alpha$, is
continuous. Therefore, to sum up, the eigenvectors and the energy spectrum of
the Hamiltonian $\hat H_{1}$ are determined by the following formulas:
\beq{\label{es17}} 
\Psi_{(n, \alpha)} &=& \Phi_{n} \otimes
\phi_{\alpha} \equiv |n,\alpha\rangle, \cr \cr \mathcal
E_{(n,\alpha)} &=& \frac{\hbar \omega_{c}}{2}(2n + 1) - \frac{\hbar
\lambda}{m}\alpha - \frac{\lambda^{2}}{2m} \quad n= 0, 1, 2,
\ldots
\eeq

\subsection{Case of the second possible symmetric gauge}

We consider now the symmetric gauge
\beq{\label{ei18}} \overrightarrow{A} = \left(-\frac{B}{2}y,
\frac{B}{2}x \right), 
\eeq 
with the scalar potential  given by \beq
\Phi(x,y) = -E x.  
\eeq 
The classical Hamiltonian $H$ in equation  (\ref{es0}) becomes
\beq{\label{ei19}}
H_{2}(x,y,p_x,p_y)  =
\frac{1}{2m}\left[\left(p_{x} - \frac{eB}{2c}y\right)^{2} +
\left(p_{y} + \frac{eB}{2c}x\right)^{2}\right] + eEx.
\eeq 
By mean of canonical quantization and proceeding like in the previous section, we define the two sets of annihilation and creation operators defined by
\beq{\label{ei20}}
\mathfrak b^{\dag} = -2i P_{\bar{z}} + \frac{eB}{2c}Z + \lambda, \qquad  \mathfrak b = 2 i P_{z} + \frac{eB}{2c}\bar{Z} + \lambda, 
\eeq 
\beq{\label{ei21}} \mathfrak d = 2i P_{z} - \frac{eB}{2c}\bar{Z},\qquad  \mathfrak d^{\dag} = -2i P_{\bar{z}} - \frac{eB}{2c}Z, 
\eeq 
with
$\lambda$ defined as in (\ref{es2}) and (\ref{es3}). They also  commute  each with other and satisfy the commutation relations (\ref{es4}). The corresponding Hamiltonian operator $\hat H_{2}$  can be then written as
\beq{\label{ei24}} 
\hat H_{2} =
\frac{1}{4m}(\mathfrak b^{\dag}\mathfrak b + \mathfrak b\mathfrak b^{\dag}) - \frac{\lambda}{2m}(\mathfrak d^{\dag} +
\mathfrak d) - \frac{\lambda^{2}}{2m}, 
\eeq 
where the following relation
\beq{\label{ei27}} \mathfrak d^{\dag} + \mathfrak d = 2P_{y} - \frac{eB}{c}X,
\eeq is obtained. Here, the harmonic oscillator part is given by 
\beq
\hat H_{2_{OSC}} = \frac{1}{4m}(\mathfrak b^{\dag}\mathfrak b + \mathfrak b\mathfrak b^{\dag})
\eeq
and the linear part by 
\beq
\hat T_{2} = \frac{\lambda}{2m}(\mathfrak d^{\dag} +
\mathfrak d) + \frac{\lambda^{2}}{2m}.
\eeq

The annihilation and creation operators $\mathfrak b$ and $\mathfrak b^{\dag}$ become
here \beq \mathfrak b = \sqrt{2m\hbar \omega_{c}}\mathfrak b',\quad 
\mathfrak b^{\dag} =
\sqrt{2m\hbar \omega_{c}}\mathfrak b'^{\dag}, \quad [\mathfrak b',\mathfrak b'^{\dag}] =
\1, 
\eeq with
\beq{\label{boson1}} \mathfrak b' =
\sqrt{\frac{2m\omega_{c}}{\hbar}}\left(\frac{i
P_{z}}{m\omega_{c}} + \frac{\bar{Z}}{4} +
\frac{\lambda}{2m\omega_{c}}\right), \quad \mathfrak b'^{\dag} = \sqrt{\frac{2m\omega_{c}}{\hbar}}\left(-\frac{i
P_{\bar{z}}}{m\omega_{c}} + \frac{Z}{4} +
\frac{\lambda}{2m\omega_{c}}\right).
\eeq 
From (\ref{ei27}), it comes \beq{\label{ei28}}
\frac{\lambda}{2m}(\mathfrak d^{\dag} + \mathfrak d)  =  \frac{\hbar
\lambda}{m}\left(\frac{P_{y}}{\hbar} - \frac{1}{2}\frac{m
\omega_{c}}{\hbar} X \right) = 
\frac{\hbar \lambda}{m}
\left(-i\frac{\partial}{\partial y} - \frac{m \omega_{c}}{2 \hbar}X \right).
\eeq 
Then, the eigenvalue equation $\hat T_2\phi = \mathcal E \phi$ is equivalent in this case to 
\beq \frac{\hbar \lambda}{m}\left(-i
\frac{\partial}{\partial y} - \frac{m \omega_{c}}{2 \hbar}X
\right)\phi = \mathcal E \phi, \eeq which leads to \beq
\left(-i \frac{\partial}{\partial y} - \frac{m \omega_{c}}{2
\hbar}x \right)\phi - 
\frac{m \mathcal E}{\hbar \lambda}\phi = 0.
\eeq 
Taking again $\alpha = \frac{m \mathcal E}{\hbar \lambda}$, it follows the equation
\beq -i \frac{\partial}{\partial y}\phi =
\left(\frac{m \omega_{c}}{2 \hbar}x + \alpha\right)\phi, \eeq which can be solved to give the eigenfunctions \beq{\label{vep2}}
\phi_{\alpha} \equiv \phi_{\alpha}(x,y) = e^{i (\alpha y +
\frac{m \omega_{c}}{2 \hbar}xy)} \;\;\; \alpha \in \R,
\eeq 
of the operator  $\hat T_2$ corresponding to eigenvalues expressed  as in (\ref{es16}). Therefore, the eigenvectors and eigenvalues of the Hamiltonian $\hat H_{2}$, as previously determined for $\hat H_{1}$, are obtained  as 
\beq{\label{eig003}} 
\Psi_{(l, \alpha)} &=& \Phi_{l} \otimes
\phi_{\alpha} \equiv |l,\alpha\rangle, \cr \cr \mathcal
E_{(l,\alpha)} &=& \frac{\hbar \omega_{c}}{2}(2l + 1) - \frac{\hbar\lambda}{m}\alpha - \frac{\lambda^{2}}{2m} \;\;\; \;\; l= 0, 1, 2,
\ldots 
\eeq
Introduce the position and momentum operators obtained from the annihilation and creation operators (\ref{es2}) and (\ref{ei20}) as 
\beq
\hat Q_1  &=& \frac{1}{2\sqrt{m\omega_c \hbar}}(b^{\dag}+ b), \quad \hat P_1  = \frac{i}{2\sqrt{m\omega_c \hbar}}(b^{\dag}- b),\cr
\hat Q_2 &=& \frac{1}{2\sqrt{m\omega_c \hbar}}(\mathfrak b^{\dag}+ \mathfrak b), \quad \hat P_2  = \frac{i}{2\sqrt{m\omega_c \hbar}}(\mathfrak b^{\dag}- \mathfrak b), 
\eeq
respectively, where the following commutation relations
\beq
[b, \mathfrak{b}^{\dag}] &=& 0 = [b^{\dag}, \mathfrak{b}], \quad [b, \mathfrak{b}] = 0 = [b^{\dag}, \mathfrak{b}^{\dag}], \cr
[\hat Q_{1},\hat P_{2}] &=& 0= [\hat Q_{2}, \hat P_{1}],\quad [\hat Q_{1}, \hat Q_{2}] = 0 = 
[\hat P_{1}, \hat P_{2}]
\eeq
are satisfied. 
Then, we respectively  have in the gauges $\overrightarrow{A} = \left(\frac{B}{2}y,
-\frac{B}{2}x \right)$  and $\overrightarrow{A} = \left(-\frac{B}{2}y,
\frac{B}{2}x \right)$ 
\beq\label{hacom00}
\hat H_{1_{OSC}} = \frac{\hbar
\omega_{c}}{2}[{Q}^{2}_{1} + {P}^{2}_{1}], \quad 
\hat H_{2_{OSC}}
= \frac{\hbar \omega_{c}}{2}[{Q}^{2}_{2} + {P}^{2}_{2}], \quad [\hat H_{1_{OSC}}, \hat H_{2_{OSC}}] = 0.
\eeq
Thus, from (\ref{es17}), (\ref{eig003}) and (\ref{hacom00}), the eigenvectors denoted $|\Psi_{nl}\rangle := |n, l\rangle = |n\rangle \otimes |l\rangle$ of $\hat H_{1_{OSC}} $ can be so chosen that they are also
the eigenvectors of $\hat H_{2_{OSC}}$ as follows:
\beq{\label{equa37}}
\hat H_{1_{OSC}}|\Psi_{nl}\rangle  = \hbar \omega_{c} \left(n + \frac{1}{2}\right)
|\Psi_{nl}\rangle, \,\,  \hat H_{2_{OSC}}|\Psi_{nl}\rangle  = \hbar
\omega_{c} \left(l + \frac{1}{2}\right) |\Psi_{nl}\rangle, \; n, l  = 0,1,2,\dots, \infty
\eeq
so that $\hat H_{2_{OSC}}$ lifts the degeneracy of $\hat H_{1_{OSC}}$ and vice versa.

From (\ref{es16}), 
consider the shifted eigenvalues
\beq\label{contvapshif}
\mathcal E'_{\alpha} :=  \mathcal E_{\alpha} -
\frac{\lambda^{2}}{2m} = \frac{\hbar \lambda}{m}\alpha, 
\eeq 
where the
states  $|\epsilon_{\alpha}\rangle$ are delta-normalized states and form the orthonormal basis $\{|\epsilon_{\alpha}\rangle, \alpha \in \R \}$. The satisfy the eigenvalue equation
\beq\label{teignefunc}
\left(\hat T_1 - \frac{\lambda^{2}}{2m}  I_{\mathfrak
H_{C}}\right)|\epsilon_{\alpha}\rangle = \mathcal E'_{\alpha}|\epsilon_{\alpha}\rangle\;,
\eeq
 which is the same equation for the operator $\hat T_2$.
 
\section{Construction of coherent states}\label{sec3}

In this section, CSs are constructed, considering the two possible orientations  of the magnetic field as
 in \cite{ab1} as well as additional parameters, originated from discrete and continuous aspects of the Hamiltonian spectrum in line with \cite{gazeau-klauder}. As a matter of comparison, we first replace the original Hamiltonian operators by their corresponding shifted counterparts, as done in \cite{gazeau-klauder}. Then, we investigate the full operators and analyze the results.

\subsection{Case of the shifted quantum Hamiltonian}\label{firstconst}

Let  $\mathfrak{H}_{D+C}:=\mathfrak{H}_D \oplus \mathfrak{H}_C$ be the Hilbert space associated to the operator $\mathcal{H}_{D} \oplus \mathcal{H}_{C}$, where $\mathfrak{H}_D$ and $\mathfrak{H}_C$ are associated to discrete and continuous spectra, respectively.  Let consider the discrete shifted Hamiltonian  ${\mathcal H}_{D} := {H}_{1_{osc}} -  \frac{\hbar \omega_{c}}{2}  \1_{\mathfrak H_{D}}$ and the continuous shifted   Hamiltonian ${\mathcal H}_{C} := T_{1} - \frac{\lambda^{2}}{2m} \1_{\mathfrak H_{C}}$, where $\1_{\mathfrak H_{D}}$ and $\1_{\mathfrak H_{C}}$ denote the identity operators on $\mathfrak{H}_D$ and $\mathfrak{H}_C$, respectively. 
 Let  $\mathfrak{H}_D$  spanned by the eigenvectors 
 $|\Psi_{nl}\rangle \equiv |n,l\rangle$ of $H_{1_{OSC}}$ and $H_{2_{OSC}}$ provided by (\ref{equa37}).  Besides, let $\mathfrak{H}_C$ be  the Hilbert space associated to the continuous spectrum spanned by the eigenvectors of the operator  $T_1$  denoted $|\epsilon_{\alpha}\rangle$ in  equation (\ref{teignefunc}).
 
The shifted Hamiltonian $\left({H}_{1_{osc}} -  \frac{\hbar \omega_{c}}{2}  \1_{\mathfrak H_{D}} \right)-
\left(T_{1} - \frac{\lambda^{2}}{2m} \1_{\mathfrak H_{C}}\right)$ possesses a spectrum which is discrete and
degenerate according to (\ref{es17}); the Landau levels are infinitely degenerate and given by
$\{\mathcal E'_{n_{osc}} = \hbar
\omega_{c} n, n = 0, 1, 2, \dots\}$ while the continuous spectrum is furnished by 
$\{ \mathcal E'_{\alpha}, \alpha \in \R\}$.
So, from (\ref{tfunc1}) and (\ref{oscil1}), 
the positive eigenvalues are 
\beq\label{tfunc00}
\mathcal E'_{n,\alpha} = \mathcal E'_{n} - \mathcal E'_{\alpha} =
\hbar \omega_{c} \left(n - \frac{\lambda}{m\omega_{c}}\alpha \right) = \hbar \omega_{c} \left(n - \epsilon_{\alpha}\right), \; \epsilon_{\alpha} = \frac{\lambda}{m\omega_{c}}\alpha,
\eeq
such that,  for all $n \in \N^{*}$, \, $\alpha \leq \frac{m \omega_{c}}{\lambda}$.
For the continuous spectrum, one also requires  the condition  $\mathcal E'_{\alpha} = -\hbar \omega_{c} \epsilon_{\alpha} \geq 0$, which implies  $\alpha \leq 0$. Therefore, the energy positivity condition should be:
$\alpha \leq 0$.

Provided the positivity of the eigenvalues, as required  \cite{ab1}
for the operator $T_{1}$ (respectively $T_{2}$),
the CSs related to  ${\mathcal H}_{D} \oplus {\mathcal H}_{C}$, are given by the
{\it unnormalized} states \cite{gazeau-klauder}
\beq{\label{el1}}
|J,\gamma;J',\gamma';l;K,\theta;\beta\rangle
&=& f(K,\theta)|J,\gamma;J',\gamma';l\rangle  + e^{-i \beta}g(J, \gamma,J',\gamma') |K, \theta\rangle \cr
&=& f(K,\theta)\left[\mathcal N(J) \mathcal N(J')\right]^{-1/2}J'^{l/2}e^{i l \gamma'}\sum^{\infty}_{n=0}\frac{J^{n/2}e^{-i n \gamma}}{\sqrt{n !l !}}|\Psi_{nl}\rangle  \cr
&& + e^{-i \beta}g(J,\gamma,J',\gamma')\mathcal N_{\rho}(K)^{-1/2}\int^{\infty}_{0}\frac{K^{\epsilon^{-}_{\alpha}/2}e^{i \epsilon_{\alpha}\theta}}{\sqrt{\rho(\epsilon^{-}_{\alpha})}} |\epsilon^{-}_{\alpha}\rangle d\epsilon^{-}_{\alpha},
\eeq
with $\epsilon^{-}_{\alpha} =: - \epsilon_{\alpha} \geq 0$. The labeling parameters are chosen such that: $ 0\leq J,J',K \leq \infty,\;  -\infty < \gamma, \gamma', \theta < \infty $ and $0 \leq \beta < 2\pi$. $f$ and $g$ are scalar functions and the normalization constants are given by
with $|J,\gamma;J',\gamma';l\rangle \in \mathfrak{H}_D$ and 
\beq{\label{el6}}
\sum^{\infty}_{l=0}\langle J,\gamma;J',\gamma';l|J,\gamma;J',\gamma';l\rangle  = \frac{1}{\mathcal N(J)}\sum^{\infty}_{n=0}\frac{J^{n}}{n !}\frac{1}{\mathcal N(J')}
\sum^{\infty}_{l=0}\frac{J'^{l}}{l !} = 1.
\eeq
Besides, $|K,\theta\rangle \in \mathfrak{H}_C$ and
\beq{\label{el7}}
\langle K,\theta|K,\theta\rangle  &=& 1 \Rightarrow \mathcal N_{\rho}(K)^{-1}\int^{\infty}_{0}
\frac{K^{\epsilon^{-}_{\alpha}}}{\rho(\epsilon^{-}_{\alpha})}d\epsilon^{-}_{\alpha} \langle \epsilon^{-}_{\alpha}|\epsilon^{-}_{\alpha}\rangle
= \mathcal N_{\rho}(K)^{-1}\int^{\infty}_{0}\frac{K^{\epsilon^{-}_{\alpha}}}{\rho(\epsilon^{-}_{\alpha})}d\epsilon^{-}_{\alpha} \cr
\cr
&& \Rightarrow \mathcal N_{\rho}(K) = \int^{\infty}_{0}\frac{K^{\epsilon^{-}_{\alpha}}}{\rho(\epsilon^{-}_{\alpha})}
d\epsilon^{-}_{\alpha}.
\eeq
The continuity of the combined CSs follows from the continuity of the separate states
and of that of the functions $f$ and $g$, which are assumed. Indeed, from the definition, we have
\beq
&&||\,\,  |J,\gamma;J',\gamma';l;K,\theta;\beta\rangle- |\tilde J,\tilde \gamma;\tilde{J'}, \tilde{\gamma'};l;\tilde K, \tilde \theta;\beta\rangle\, \,  ||^2 \cr
&=& |f(K,\theta)|^2 \langle J,\gamma;J',\gamma';l|J,\gamma;J',\gamma';l \rangle + 
|g(J,\gamma;J',\gamma')|^2 \langle K,\theta|K,\theta \rangle \cr
&&+|f(\tilde K,\tilde \theta)|^2 \langle \tilde J,\tilde \gamma;J',\tilde \gamma';l|\tilde J,\tilde \gamma;\tilde J',\tilde \gamma';l \rangle + 
|g(J,\gamma;J',\gamma')|^2 \langle \tilde K,\tilde \theta|\tilde K,\tilde \theta \rangle\cr
&&+ 
f(K,\theta)f(\tilde K,\tilde \theta)^{*}
\langle \tilde J,\tilde \gamma;\tilde J',\tilde \gamma';l |J,\gamma;J',\gamma';l \rangle + 
f(K,\theta)^{*}f(\tilde K,\tilde \theta)
\langle J,\gamma;J',\gamma';l|\tilde J,\tilde \gamma;\tilde J',\tilde \gamma';l \rangle\cr
&&+g(J,\gamma;J',\gamma')g(\tilde J,\tilde \gamma;\tilde J',\tilde \gamma')^{*}\langle \tilde K,\tilde \theta|K,\theta \rangle + g(J,\gamma;J',\gamma')^{*}g(\tilde J,\tilde \gamma;\tilde J',\tilde \gamma')\langle K,\theta|\tilde K,\tilde \theta\rangle
\eeq
such that 
\beq
\lim_{(J,\gamma;J',\gamma';K,\theta)
\rightarrow (\tilde J,\tilde \gamma;\tilde J',\tilde \gamma';\tilde K,\tilde \theta)}||\, \, |J,\gamma;J',\gamma';l;K,\theta;\beta\rangle - |\tilde J,\tilde \gamma;\tilde{J'}, \tilde{\gamma'};l;\tilde K, \tilde \theta;\beta\rangle \,\,  ||^2 
=0.
\eeq
Now, let us investigate the resolution of the identity or the completeness relation which is  expressed in terms of the projectors onto the states $|J,\gamma;J',\gamma';l;K,\theta;\beta\rangle$.
\bpro\label{prop1}
The CSs (\ref{el1}) satisfy, on $\mathfrak H_{D+C}$,
the resolution of the identity
\beq{\label{el3}}
&&\int^{\infty}_{0}\int^{\infty}_{0}\int^{\infty}_{0}\int_{\R}
\int_{\R}\int_{\R}\int^{2\pi}_{0}|J,\gamma;J',\gamma';l;K,\theta;\beta\rangle \langle J,\gamma;J',\gamma';l;K,\theta;\beta| \cr
&& d\mu_{B}(\gamma)d\mu_{B}(\gamma')\frac{d\theta}{2\pi}\frac{d\beta}{2\pi}\mathcal N(J)
\mathcal N(J')\mathcal N_{\rho}(K)d\nu(J)d\nu(J')d\lambda(K) = 
\1_{{\mathfrak H}^l_{D}}
+  \1_{\mathfrak H_{C}}
\eeq
where $\1_{\mathfrak H^l_D}, \1_{\mathfrak H^n_D}$  are the identity operators on the subspaces $\mathfrak H^n_D, \mathfrak H^l_D$ of $\mathfrak{H}_D$ such that 
\beq\label{identsubs}
\sum_{n=0}^{\infty}|\Psi_{nl}\rangle \langle \Psi_{nl}| = \1_{\mathfrak H^l_D}, \quad \sum_{l=0}^{\infty}|\Psi_{nl}\rangle \langle \Psi_{nl}| = \1_{\mathfrak H^n_D}.
\eeq
$d\mu_{B}$ refers to the {\it Bohr measure} \cite{ab1} provided as follows
\beq
\langle f|g\rangle_{ns} = \lim_{T \to \infty}\frac{1}{2T}\int_{-T}^{T} \overline{f(\gamma)}g(\gamma)d\gamma:= 
\int_{\R}\overline{f(\gamma)}g(\gamma)d\mu_{B}(\gamma)
\eeq
given on the Hilbert space 
$\mathfrak H_{ns}$  of functions $f : \R \rightarrow \C$, which is complete with respect to the scalar 
product $\langle .|.\rangle_{ns}$.  $d\lambda(K) = \sigma(K)dK$, and $\sigma(K)$ is a non-negative weight function $\sigma(K) \geq 0$ such that
\beq
\int_{0}^{\infty}K^{\epsilon^{-}_{\alpha}} \sigma(K) dK \equiv \rho(\epsilon^{-}_{\alpha}).
\eeq
On the Hilbert spaces  ${\mathfrak H}_{D}$, ${\mathfrak H}_{C}$ and  ${\mathfrak H}_{D+C}$,  we have the following  essential relations
\beq\label{identrq000}
&&\int_{\R}\int_{\R}\int_{\R}\int^{\infty}_{0}\int^{\infty}_{0}\int^{\infty}_{0}
\int^{2\pi}_{0}|f(K,\theta)|^{2}|J,\gamma;J',\gamma';l \rangle \langle J,\gamma;J',\gamma';l| \cr
&&d\mu_{D}(J,\gamma,J',\gamma')d\mu_{C}(K,\theta)\frac{d\beta}{2\pi} = I_{\mathfrak H^l_{D}}, \cr
\cr
&&\int_{\R}\int_{\R}\int_{\R}\int^{\infty}_{0}\int^{\infty}_{0}\int^{\infty}_{0}
\int^{2\pi}_{0}|g(J,\gamma,J',\gamma')|^{2}|K,\theta\rangle \langle K,\theta| \cr
&&d\mu_{D}(J,\gamma,J',\gamma')d\mu_{C}(K,\theta)\frac{d\beta}{2\pi} = \1_{{\mathfrak H}_{C}}, \cr
\cr
&&\int^{2\pi}_{0}e^{i \beta}\frac{d\beta}{2\pi}\int_{\R}\int_{\R}\int_{\R}\int^{\infty}_{0}\int^{\infty}_{0}
\int^{\infty}_{0}g(J,\gamma,J',\gamma')^{*}f(K,\theta)
|J,\gamma;J',\gamma';l \rangle \langle K,\theta| \cr
&&d\mu_{D}(J,\gamma,J',\gamma')d\mu_{C}(K,\theta) = 0.
\eeq
that need to be satisfied, where $d\mu_{D}$ and $d\mu_{C}$ are the measures associated to the discrete-spectrum CSs $\{J,\gamma,J',\gamma'\}$ and continuous-spectrum CSs $\{K,\theta\}$ labeling parameters, respectively. 
The identity operator $\1_{\mathfrak H_{D+C}}$ is the direct sum of
the identity operators $\1_{{\mathfrak H}_{D}}$ and $\1_{{\mathfrak H}_{C}}$ which act
on the complementary subspaces ${\mathfrak H}_{D}$ and ${\mathfrak H}_{C}$, respectively, corresponding to discrete  and continuous spectra.
\epro
Noting that the integration over $\beta, \, 0 \leq \beta < 2\pi$ eliminates the third relation above, which is related to the off-diagonal terms, the three conditions (\ref{identrq000}) are reduced to
\beq{\label{el4}}
&&\int_{\R}\int^{\infty}_{0}
|f(K,\theta)|^{2}d\mu_{C}(K,\theta) = 1,
\cr
&&\int_{\R}\int_{\R}\int^{\infty}_{0}\int^{\infty}_{0}|g(J,\gamma,J',\gamma')|^{2} d\mu_{D}(J,\gamma,J',\gamma') = 1.
\eeq
In view of getting the resolution of the identity, let us take the functions $f$ and $g$ as in \cite{gazeau-klauder}, such that
\beq
f(K,\theta) = \mathcal N_{f} \, e^{-\frac{K^{2} + \theta^{2}}{2}}, \;\;\;
g(J,\gamma,J',\gamma') = \mathcal N_{g} \, e^{-\frac{J^{2} + J'^{2}}{2}},
\eeq
where the factors $\mathcal N_{g}$ and $\mathcal N_{f}$ are chosen so that
\beq
\mathcal N^{2}_{f}\int_{\R}\int_{0}^{\infty}
e^{-(K^{2} + \theta^{2})}
d\mu_{C}(K,\theta) = 1, \cr
\cr
\mathcal N^{2}_{g}\int_{\R}\int_{\R}\int_{0}^{\infty}\int_{0}^{\infty}e^{-(J^{2} + J'^{2})}
d\mu_{D}(J,\gamma,J',\gamma') = 1.
\eeq
{\bf Proof.} See in the Appendix \ref{app000}.
$\hfill{\square}$
\bpro\label{prop2}
The property of temporal stability can be obtained here by postulating similar assumptions as  in \cite{gazeau-klauder},
such that $0 \leq \mathcal H_{D} \leq \Omega$ and $\Omega < \mathcal H_{C}$, i.e. the Hamiltonians are adjusted so that $0< \mathcal H_{C} - \Omega$. By taking into account the phase factor $e^{-i \beta }$, it comes the following relation
\beq
e^{-i \mathcal H t}|J,\gamma;J',\gamma';l;K,\theta;\beta\rangle 
&=& f(K,\theta)|
J,\gamma + \omega_{c} t;J',\gamma';l\rangle \cr
&&+ e^{-i (\beta + \Omega t)} g(J,\gamma,J',\gamma')|K, \theta + \omega_{c} t\rangle \cr
&=& |J,\gamma + \omega_{c} t;J',\gamma';l;K,\theta + \omega_{c} t;\beta + \Omega t\rangle,
\eeq
with $\mathcal H = \mathcal H_{D} + (\mathcal H_{C} - \Omega)$.
\epro
{\bf Proof.}  See in the Appendix \ref{app000}.
$\hfill{\square}$

The action identity as noticed in \cite{gazeau-klauder} is difficult to obtain with the combined CSs given in 
 (\ref{el1}).
 
\subsection{Case of the  Hamiltonian ${H}_{2_{osc}} - T_{2}$}

By analogy of the setting in Section \ref{firstconst}, we study the shifted Hamiltonian $\left({H}_{2_{osc}} -  \frac{\hbar \omega_{c}}{2}  \1_{\mathfrak H_{D}} \right)- \left(T_{2} - \frac{\lambda^{2}}{2m} \1_{\mathfrak H_{C}}\right)$.

The related CSs are here given on $\mathfrak H_{D+C}$ by
\beq{\label{el5}}
|J,\gamma;J',\gamma';n;K,\theta;\beta\rangle
&=& f(K,\theta)|J,\gamma;J',\gamma';n\rangle + e^{-i \beta}g(J, \gamma,J',\gamma') |K, \theta\rangle \cr
&=&f(K,\theta)\left[\mathcal N(J) \mathcal N(J')\right]^{-1/2}J^{n/2}e^{-i n \gamma}\sum^{\infty}_{l=0}\frac{J'^{l/2}e^{i l \gamma'}}{\sqrt{n !l !}}|\Psi_{nl}\rangle  \cr
&& + e^{-i \beta}g(J,\gamma,J',\gamma')\mathcal N_{\rho}(K)^{-1/2}\int^{\infty}_{0}\frac{K^{\epsilon^{-}_{\alpha}/2}e^{i \epsilon_{\alpha}\theta}}{\sqrt{\rho(\epsilon^{-}_{\alpha})}}|\epsilon^{-}_{\alpha}\rangle d\epsilon^{-}_{\alpha},
\eeq
where the normalization constants are given as  in (\ref{el6})
with the relation (\ref{el7}) also satisfied.
\bpro
The CSs satisfy, on $\mathfrak H_{D+C}$, the resolution of the identity
\beq{\label{el9}}
&&\int^{\infty}_{0}\int^{\infty}_{0}\int^{\infty}_{0}\int_{\R}
\int_{\R}\int_{\R}\int^{2\pi}_{0}|J,\gamma;J',\gamma';n;K,\theta;\beta\rangle \langle J,\gamma;J',\gamma';n;K,\theta;\beta| \cr
&& d\mu_{B}(\gamma)d\mu_{B}(\gamma')\frac{d\theta}{2\pi}\frac{d\beta}{2\pi}\mathcal N(J)
\mathcal N(J')\mathcal N_{\rho}(K)d\nu(J)d\nu(J')d\lambda(K) = 
\1_{{\mathfrak H}^n_{D}}
+  \1_{\mathfrak H_{C}}.
\eeq
\epro
{\bf Proof.} See that of Proposition \ref{prop1}.
$\hfill{\square}$
\bpro
The temporal stability property is given by
\beq
e^{-i \mathcal H t}|J,\gamma;J',\gamma';n;K,\theta;\beta\rangle &=& f(K,\theta)|
J,\gamma;J',\gamma' + \omega_{c} t;n\rangle \cr
&&+ e^{-i (\beta + \Omega t)} g(J,\gamma,J',\gamma')|K, \theta + \omega_{c} t\rangle \cr
&=& |J,\gamma;J',\gamma' + \omega_{c} t;n;K,\theta + \omega_{c} t;\beta + \Omega t\rangle.
\eeq
\epro
{\bf Proof.} See that of Proposition \ref{prop2}.
$\hfill{\square}$

\subsection{Case of the unshifted Hamiltonians $H_{1}$ and $H_{2}$}

The eigenvalues $\mathcal E_{n,\alpha}$ of the Hamiltonian operators $H_{1}$ and $H_{2}$, given respectively in Eq.(\ref{es17}) and (\ref{eig003}), can be rewritten as $\mathcal E_{n,\alpha} = \mathcal E_{n} + \mathcal E_{\alpha}$, where
\beq
\mathcal E_{n} = \hbar \omega_{c} \left(n+ \frac{1}{2}\right), \; \mathcal E_{\alpha} = -\hbar \omega_{c} \epsilon_{\alpha}, \; \epsilon_{\alpha} = \frac{\lambda}{m \omega_{c}}\alpha + \frac{\lambda^{2}}{2m\hbar \omega_{c}}.
\eeq
The required conditions $\mathcal E_{n,\alpha} \geq 0$ for all $n \in \N$ and $\mathcal E_{\alpha} \geq 0$ lead to the relations
\beq
\alpha \leq \frac{ m\omega_{c}}{2\lambda} - \frac{\lambda}{2\hbar}, \; \alpha \leq -\frac{\lambda}{2\hbar},
\eeq
giving $\alpha \leq -\frac{\lambda}{2\hbar}$. Setting
\beq
\rho(n) = \mathcal E_{1}\mathcal E_{2}\dots \mathcal E_{n}
\eeq
such that
\beq\label{conststruc}
\rho(n) = \prod_{k=1}^{n}\hbar \omega_{c}\left(k+\frac{1}{2}\right)  = (\kappa)^{n}\left(\frac{3}{2} \right)_{n}, \; \kappa = \hbar \omega_{c}
\eeq
where $(\frac{3}{2})_{n}$ stands for the Pochhammer symbol \cite{erdelyi-ismail}. 
  The CSs,    related to $H_{1}$, defined in line with (\ref{el1}), are now given,  
%   as in (\ref{el1})
  where the function $\rho(n)= n !$ is replaced by the one in (\ref{conststruc}), 
  by
\beq{\label{es03}}
|J,\gamma;J',\gamma';l;K,\theta;\beta\rangle
&=& f(K,\theta)\left[\mathcal N(J) \mathcal N(J')\right]^{-1/2}J'^{l/2}e^{i \mathcal E_{l} \gamma'}\sum^{\infty}_{n=0}\frac{J^{n/2}e^{-i \mathcal E_{n} \gamma}}{\sqrt{\rho(n)\rho(l)}}|\Psi_{nl}\rangle  \cr
&& + e^{-i \beta}g(J,\gamma,J',\gamma')\mathcal N_{\rho}(K)^{-1/2}\int^{\infty}_{0}\frac{K^{\epsilon^{-}_{\alpha}/2}e^{i \epsilon_{\alpha}\theta}}{\sqrt{\rho(\epsilon^{-}_{\alpha})}} |\epsilon^{-}_{\alpha}\rangle d\epsilon^{-}_{\alpha}
\eeq
yields
\beq
\mathcal N(J) &=& \sum_{n=0}^{\infty}\frac{J^{n}}{\rho(n)} = \sum_{n=0}^{\infty}\frac{J^{n}}{\kappa^{n}(\frac{3}{2})_{n}} = \,  _{1}F_{1}\left(1;\frac{3}{2};\frac{J}{\kappa}\right), \cr
\cr
\mathcal N(J') &=& \sum_{l=0}^{\infty}\frac{J'^{l}}{\rho(l)} =
\sum_{l=0}^{\infty}\frac{J'^{l}}{\kappa^{l}(\frac{3}{2})_{l}} = \,  _{1}F_{1}\left(1;\frac{3}{2};\frac{J'}{\kappa}\right),
\eeq
with the relation (\ref{el7}) also remaining here valid.
\bpro
The CSs (\ref{es03}) satisfy, on $\mathfrak H_{D+C}$, a resolution of the identity  given in (\ref{el3}), where the measures $d\nu(J)$ and $d\nu(J')$ are  now given by
\beq
d\nu(J) &=& \frac{\Gamma^{2}(n+1)}{\kappa^{\eta}\Gamma(\eta)}   \, _{1}F_{1}\left(1;\eta;\frac{J}{\kappa}\right) 
\frac{e^{-J/\kappa}J^{-n + \eta - 1}}{(\frac{1}{\kappa} + \mu - \sigma)^{n}}
L^{\eta - 1}_{n}[(\mu - \sigma)J]dJ,  \cr
\cr
d\nu(J')&=& \frac{\Gamma^{2}(l+1)}{\kappa^{\eta}\Gamma(\eta)}   \, _{1}F_{1}\left(1;\eta;\frac{J'}{\kappa}\right) 
\frac{e^{-J'/\kappa}J'^{-l + \eta - 1}}{(\frac{1}{\kappa} + \mu - \sigma)^{l}}
L^{\eta - 1}_{l}[(\mu - \sigma)J']dJ', \, \eta = \frac{3}{2},
\eeq
where the quantities $L^{\eta - 1}_{n}[(\mu - \sigma)J], L^{\eta - 1}_{l}[(\mu - \sigma)J']$ are the  Laguerre polynomials,  and lead to the identities \cite{erdelyi-ismail}
\beq
n !\int_{0}^{\infty}t^{\nu-n}e^{\mu t} L_{n}^{\nu-n}[(\mu - \sigma)t]e^{-st}dt 
=\Gamma(\nu+1)(s-\sigma)^{n}(s-\mu)^{-\nu-1}, \; \Re{(\nu)} > n-1 \crcr
l !\int_{0}^{\infty}t^{\nu-l}e^{\mu t} L_{l}^{\nu-l}[(\mu - \sigma)t]e^{-st}dt 
=\Gamma(\nu+1)(s-\sigma)^{l}(s-\mu)^{-\nu-1}, \; \Re{(\nu)} > l-1
\eeq 
with $\nu = n+\eta-1$ (resp. $\nu = l + \eta-1$) and $\frac{1}{\kappa} = s-\mu$.
\epro
 The CSs for the Hamiltonian $H_{2}$,  similar to the ones in (\ref{es03}),  can be  constructed in the same way, with the labeling parameters $J,\gamma$ playing  the role of
$J',\gamma'$ and vice versa.

\section{Concluding remarks}\label{sec4}

Coherent states have been constructed  for Hamiltonians with both discrete and continuous spectra, in the context of the motion of an electron in an electromagnetic field, arising in the  quantum Hall effect by considering shifted and unshifted spectra, respectively.
 These coherent states satisfy the  Gazeau-Klauder coherent states criteria that are the  continuity in the labels, the resolution of the identity and the  temporal stability. The action identity property remains difficult to  obtain  in the combined coherent states as noticed in \cite{gazeau-klauder}.
 
 An extension of this work that is currently under investigation is the construction of coherent states for an  Hamiltonian in the case of an  electric field depending simultaneously on  both $x$ and $y$ directions, and for Hamiltonian operators  admitting discrete eigenvalues and eigenfuctions in appropriate  Hilbert space \cite{aremuaetal}.

\section{Appendix}\label{app000}

{\bf Proof of Proposition \ref{prop1}} From (\ref{el1}), we get
\beq
&&|J,\gamma;J',\gamma';l;K,\theta;\beta\rangle \langle J,\gamma;J',\gamma';l;K,\theta;\beta| \cr
&& = |f(K,\theta)|^{2}\frac{J'^{l}}{\left[\mathcal N(J) \mathcal N(J')\right]}\sum^{\infty}_{n,p=0}
\frac{J^{\frac{n+p}{2}}e^{i (p-n)\gamma}}{\sqrt{n !l !p !l !}}|\Psi_{nl}\rangle  \langle \Psi_{pl}|  \cr
&& + |g(J,\gamma,J',\gamma')|^{2}\mathcal N_{\rho}(K)^{-1}\int^{\infty}_{0}\int^{\infty}_{0}
\frac{K^{\frac{\epsilon^{-}_{\alpha}
+ \epsilon^{-}_{\alpha'}}{2}}e^{i (\epsilon_{\alpha} - \epsilon_{\alpha'})\theta}}
{\sqrt{\rho(\epsilon^{-}_{\alpha})\rho(\epsilon^{-}_{\alpha'})}}
|\epsilon^{-}_{\alpha}\rangle \langle \epsilon^{-}_{\alpha'}|d\epsilon^{-}_{\alpha}
d\epsilon^{-}_{\alpha'} \cr
&&+ f(K,\theta)^{*}e^{-i \beta} g(J,\gamma,J',\gamma') |K,\theta \rangle
\langle J;\gamma,J';\gamma'|
 \cr
&& + f(K,\theta)e^{i \beta}g(J,\gamma,J',\gamma')^{*}|J,\gamma;J',\gamma'\rangle \langle K,\theta|.
\eeq
Then,  the following equalities are valid:
\beq
&&\int_{\R}\int_{\R}\int_{\R}\int^{2\pi}_{0}|J,\gamma;J',\gamma';l;K,\theta;\beta\rangle
\langle J,\gamma;J',\gamma';l;K,\theta;\beta| \cr
\cr
&& d\mu_{B}(\gamma)d\mu_{B}(\gamma')\mathcal N(J)\mathcal N(J')\mathcal N_{\rho}(K)
\frac{d\theta}{2\pi}
\frac{d\beta}{2\pi} \cr
\cr
&& = \sum^{\infty}_{n,p=0}\frac{1}{\sqrt{n !l !p !l !}}\left[ \int_{\R}
|f(K,\theta)|^{2} \mathcal N_{\rho}(K)\frac{d\theta}{2\pi}\right]\int^{2\pi}_{0}\int^{2\pi}_{0}
J^{\frac{n+p}{2}}J'^{l} \cr
\cr
&& \frac{d\gamma'}{2\pi}\frac{d\beta}{2\pi}\delta_{np}|\Psi_{nl}\rangle  \langle \Psi_{pl}|  +
 \left[ \int_{\R}\int_{\R}|g(J,\gamma,J',\gamma')|^{2}
d\mu_{B}(\gamma)d\mu_{B}(\gamma')\mathcal N(J) \mathcal N(J')\right]
\cr
\cr
&& \int^{\infty}_{0}\int^{\infty}_{0}\int^{2\pi}_{0}
\frac{K^{\frac{\epsilon^{-}_{\alpha} + \epsilon^{-}_{\alpha'}}{2}}}
{\sqrt{\rho(\epsilon^{-}_{\alpha})\rho(\epsilon^{-}_{\alpha'})}}
\delta(\epsilon_{\alpha} - \epsilon_{\alpha'} ) \frac{d\beta}{2\pi}
d\epsilon^{-}_{\alpha}d\epsilon^{-}_{\alpha'}
 |\epsilon^{-}_{\alpha}\rangle \langle \epsilon^{-}_{\alpha'}| \cr
\cr
&& = \sum^{\infty}_{n=0}\frac{1}{n !l !}\left[ \int_{\R}
|f(K,\theta)|^{2} \mathcal N_{\rho}(K)\frac{d\theta}{2\pi}\right]
J^{n}J'^{l}
|\Psi_{nl}\rangle  \langle \Psi_{pl}|   \cr
\cr
&&  + \left[ \int_{\R}\int_{\R}|g(J,\gamma,J',\gamma')|^{2} d\mu_{B}(\gamma)d\mu_{B}(\gamma')
\mathcal N(J) \mathcal N(J')\right]
\int^{\infty}_{0}\frac{K^{\epsilon^{-}_{\alpha}}}{\rho(\epsilon^{-}_{\alpha})}
d\epsilon^{-}_{\alpha}
|\epsilon^{-}_{\alpha}\rangle \langle \epsilon^{-}_{\alpha}|.
\eeq
The conditions (\ref{el4}) implies
\beq
\int_{\R}\int^{\infty}_{0}
|f(K,\theta)|^{2} \mathcal N_{\rho}(K)d\lambda(K)\frac{d\theta}{2\pi} = \int_{\R}\int^{\infty}_{0}
|f(K,\theta)|^{2}d\mu_{C}(K,\theta) = 1,
\eeq
\beq
&&\int_{\R}\int_{\R}\int^{\infty}_{0}\int^{\infty}_{0}|g(J,\gamma,J',\gamma')|^{2} d\mu_{B}(\gamma)d\mu_{B}(\gamma') \mathcal N(J) \mathcal N(J')d\nu(J)d\nu(J')\cr
\cr
&&= \int_{\R}\int_{\R}\int^{\infty}_{0}\int^{\infty}_{0}|g(J,\gamma,J',\gamma')|^{2} d\mu_{D}(J,\gamma,J',\gamma') = 1.
\eeq
The measures $d\nu(J) = e^{-J}dJ$ and  $d\nu(J') = e^{-J'}dJ'$  are such that the moment problems given by
\beq
\int^{\infty}_{0}\frac{J^{n}}{n !}d\nu(J) = 1, \;\;\;
\int^{\infty}_{0}\frac{J'^{l}}{l !}d\nu(J') = 1
\eeq
are satisfied. Therefore,
\beq
&&\int^{\infty}_{0}\int^{\infty}_{0}\int^{\infty}_{0}\int_{\R}
\int_{\R}\int_{\R}\int^{2\pi}_{0}|J,\gamma;J',\gamma';l;K,\theta;\beta\rangle
\langle J,\gamma;J',\gamma';l;K,\theta;\beta| \cr
\cr
&&\times d\mu_{B}(\gamma)d\mu_{B}(\gamma')\frac{d\theta}{2\pi}\frac{d\beta}{2\pi}\mathcal N(J)
\mathcal N(J')\mathcal N_{\rho}(K)d\nu(J)d\nu(J')d\lambda(K) \cr
\cr
&& = \sum^{\infty}_{n=0}|\Psi_{nl}\rangle  \langle \Psi_{nl}|
+ \int^{\infty}_{0}|\epsilon^{-}_{\alpha}\rangle \langle \epsilon^{-}_{\alpha}|d\epsilon^{-}_{\alpha}
= I_{\mathfrak H^l_{D}} +  \1_{\mathfrak H_{C}}.
\eeq
$\hfill{\square}$
{\bf Proof of Proposition \ref{prop2}}
 By definition, we have
\beq
&& e^{-i \mathcal H t}|J,\gamma;J',\gamma';l;K,\theta;\beta\rangle \cr
&\equiv&
f(K,\theta)\left[\mathcal N(J) \mathcal N(J')\right]^{-1/2}J'^{l/2}e^{i l \gamma'}\sum^{\infty}_{n=0}\frac{J^{n/2}e^{-i n \gamma}}{\sqrt{n !l !}}e^{-i \mathcal H_{D}t}|\Psi_{nl}\rangle  \cr
&& + e^{-i (\mathcal H_{C} - \Omega)t}e^{-i \beta}g(J,\gamma,J',\gamma')\mathcal N_{\rho}(K)^{-1/2}\int^{\infty}_{0}\frac{K^{\epsilon^{-}_{\alpha}/2}e^{i \epsilon_{\alpha}\theta}}{\sqrt{\rho(\epsilon^{-}_{\alpha})}} |\epsilon^{-}_{\alpha}\rangle d\epsilon^{-}_{\alpha}
\cr
\cr
&=& f(K,\theta)\left[\mathcal N(J) \mathcal N(J')\right]^{-1/2}J'^{l/2}e^{i l \gamma'}\sum^{\infty}_{n=0}\frac{J^{n/2}e^{-i n (\gamma + \omega_{c} t)}}{\sqrt{n !l !}}|\Psi_{nl}\rangle  \cr
&& + e^{-i (\beta + \Omega t)}g(J,\gamma,J',\gamma')\mathcal N_{\rho}(K)^{-1/2}\int^{\infty}_{0}\frac{K^{\epsilon^{-}_{\alpha}/2}e^{i \epsilon_{\alpha}(\theta + \omega_{c} t)}}{\sqrt{\rho(\epsilon^{-}_{\alpha})}} |\epsilon^{-}_{\alpha}\rangle d\epsilon^{-}_{\alpha} \cr
\cr
&=& f(K,\theta)|
J,\gamma + \omega_{c} t;J',\gamma';l\rangle
+ e^{-i (\beta + \Omega t)} g(J,\gamma,J',\gamma')|K, \theta + \omega_{c} t\rangle \cr
\cr
&=& |J,\gamma + \omega_{c} t;J',\gamma';l;K,\theta + \omega_{c} t;\beta + \Omega t\rangle.
\eeq
$\hfill{\square}$
\end{document}